# The POLUSA Dataset: 0.9M Political News Articles Balanced by Time and Outlet Popularity

Lukas Gebhard
University of Freiburg, Germany
gebhardl@cs.uni-freiburg.de

Felix Hamborg
University of Konstanz, Germany
felix.hamborg@uni-konstanz.de

## ABSTRACT

News articles covering policy issues are an essential source of information in the social sciences and are also frequently used for other use cases, e.g., to train NLP language models. To derive meaningful insights from the analysis of news, large datasets are required that represent real-world distributions, e.g., with respect to the contained outlets' popularity, topically, or across time. Information on the political leanings of media publishers is often needed, e.g., to study differences in news reporting across the political spectrum, which is one of the prime use cases in the social sciences when studying media bias and related societal issues. Concerning these requirements, existing datasets have major flaws, resulting in redundant and cumbersome effort in the research community for dataset creation. To fill this gap, we present POLUSA, a dataset that represents the online media landscape as perceived by an average US news consumer. The dataset contains 0.9M articles covering policy topics published between Jan. 2017 and Aug. 2019 by 18 news outlets representing the political spectrum. Each outlet is labeled by its political leaning, which we derive using a systematic aggregation of eight data sources. The news dataset is balanced with respect to publication date and outlet popularity. POLUSA enables studying a variety of subjects, e.g., media effects and political partisanship. Due to its size, the dataset allows to utilize data-intense deep learning methods.

## 1 INTRODUCTION AND RELATED WORK

News articles serve as a crucial source of information in various disciplines, e.g., in the social sciences and digital humanities scientists analyze news coverage to understand societal issues and trends [4]. A common requirement is that the analyzed news articles reflect the news landscape, e.g., with respect to the political spectrum but also the number of readers per outlet contained in the dataset. In computer science, articles are commonly used, e.g., to train machine learning algorithms, recently – with the rise of deep learning – requiring very large amounts of data [5].

However, the creation of such datasets requires significant effort and existing news datasets suffer from at least one of the following shortcomings: (1) They *do not represent real-world distributions* of news [1–3, 7], e.g., what the average news consumer would typically read. (2) They require *considerable preprocessing effort* [5,8], e.g., because they are inefficiently accessible website dumps and contain duplicates and noise. (3) They lack an association of outlets or articles to their political leanings [1,3,5,8], a common requirement in the social sciences, or the process of deriving such associations is non-transparent [2].

To address these shortcomings, we present POLUSA, a dataset that aims to represent the landscape of online news coverage as perceived by an average US news consumer. The dataset contains 0.9M news articles covering policy topics published between Jan. 2017 and Aug. 2019 by 18 news outlets representing the political spectrum. Using a systematic aggregation of previously published measures, we label each outlet by its political slant. The dataset is balanced with respect to publication date and outlet popularity. POLUSA is available at: https://doi.org/10.5281/zenodo.3813663

## 2 DATASET CREATION

We perform five tasks, which aim to ensure the characteristics described in Section 1 and to increase the quality of the dataset: (1) base selection, (2) near-duplicate removal, (3) selection of English news articles covering policy topics, (4) assignment of political leanings, and (5) temporal and popularity-based balancing.

First, as potential news articles we extract all webpages from the commoncrawl.org news archive (CCNA) [6]. Since CCNA lacks data for various timeframes and news outlets, we need to *select a subset* of articles that is as large as possible while having an as consistent as possible number of articles for any given timeframe within. The resulting subset contains all articles of 30 news outlets, published between January 2017 and August 2019. After removing exact duplicates, the *base selection* contains 3.6M articles.

Second, we *remove near-duplicates* using nearest neighbor clustering of articles' simhashes. This way, we remove 5 % articles from the base selection, mostly consisting of outdated versions that resulted from minor article revisions, e.g., word insertions or corrections of numbers. We only keep the most recent version of an article.

Third, we *select English news articles covering policy topics*. We drop all non-English articles, which make up 6 % of the base selection. Next, we remove non-article content. To do so, we use manually derived URL heuristics. For example, the URL http://example.com/gallery/b-spears.html likely links to a gallery of photos. Using a blacklist of 47 URL segments, e.g., "/gallery/", we identify 6 % of the base selection as non-article content. To keep only articles reporting on policy topics, we use a second blacklist of 60 URL segments, e.g., "/weather/" and "/sports/". This way, we discard 13 % of the base selection. To increase the political filtering performance, we train a convolutional neural network using

GloVe word embeddings on a labeled set of 0.6M articles, extracted from CCNA and the datasets HuffPost and BBC. We fit the classifier on 87.5 % of the data. Our evaluation on 12.5 % of the data yields F1=94.4 (p=95.6, r=93.2). Finally, we only keep articles $a$ that likely report on policy topics ($p(a)≥0.75$). This applies to 49 % of the base selection. The trained classifier is available at: https://github.com/lukasgebhard/Political-News-Filter

Fourth, we *assign political leanings* to news outlets. To ensure reliable ratings, we systematically aggregate eight measures from prior work. The measures are self-declarations by outlets, results of content analyses by social scientists, and news consumer data from surveys and social networks. We label outlets with low agreement (<0.75) among the eight rating dimensions as 'undefined.' Table 1 lists the obtained political leaning ratings per outlet.

Fifth, we *balance the dataset* to decrease temporal distortions and to align it to the perceived real-world distribution. For some outlets and months, almost no or significantly fewer news articles exist in the base selection. This is likely due to technical issues during the web crawling process for CCNA. To avoid temporal distortions in the dataset, we drop ten outlets with high temporal variation. POLUSA aims to represent the landscape of online news as perceived by an average US news consumer. Therefore, each outlet's share of articles should depend on its popularity. For simplicity, we model an outlet's article count as a linear function of its popularity. We approximate an outlet's popularity by its Alexa rank. Given this model, we identify four outlets with overly many articles as compared to their rank. We randomly subsample articles of these outlets to avoid overrepresentation.

## 3 DATASET CHARACTERISTICS

POLUSA contains 0.9M news articles published by 18 outlets, including metadata such as authors, publication date, URL, as well as the publishing outlet's name and political leaning. As shown in Table 1, the largest shares of articles belong to The Guardian (11 %), Fox News (10 %) and Reuters (10 %). Left-wing outlets contribute 31 % of articles, the political center contributes 27 % and right-wing outlets contribute 16 %. 26 % of articles belong to outlets with undefined political leanings. On average, an article has about 400 to 4,000 characters, most often around 500 and 1,000 characters. Due to temporal balancing, the monthly distribution of article counts is approximately uniform across outlets. More aggressive balancing would lead to an even more uniform distribution but would strongly reduce the number of articles.

Table 1: Characteristics of outlets in POLUSA. Columns: total <u>a</u>rticle <u>c</u>ount, political <u>l</u>eaning (<u>L</u>eft, <u>C</u>enter, <u>R</u>ight), mean <u>a</u>rticle <u>l</u>ength, and the last five show the article count of the <u>k</u>-th <u>h</u>alf-year. All numbers are in units of 1K.

| Outlet | AC | PL | AL | H1 | H2 | H3 | H4 | H5 |
|---|---|---|---|---|---|---|---|---|
| ABC News | 61 | C | 3 | 17 | 11 | 14 | 12 | 6 |
| BBC | 76 | - | 3 | 3 | 12 | 20 | 17 | 19 |
| Breitbart | 57 | R | 3 | 11 | 8 | 10 | 11 | 11 |
| CBS News | 28 | C | 4 | 1 | 4 | 8 | 7 | 6 |
| CNN | 51 | - | 4 | 16 | 14 | 16 | 2 | 3 |
| Chicago Trib. | 46 | - | 4 | 10 | 8 | 8 | 9 | 10 |
| Fox News | 88 | R | 3 | 6 | 21 | 22 | 16 | 18 |
| HuffPost | 49 | L | 5 | 3 | 22 | 8 | 6 | 7 |
| LA Times | 36 | L | 3 | 9 | 8 | 5 | 6 | 5 |
| NBC News | 37 | C | 4 | 6 | 6 | 6 | 8 | 8 |
| NPR | 42 | L | 4 | 7 | 7 | 9 | 8 | 8 |
| Politico | 23 | - | 7 | 5 | 4 | 6 | 4 | 2 |
| Reuters | 87 | C | 2 | 19 | 18 | 19 | 13 | 14 |
| Slate | 15 | L | 6 | 3 | 3 | 3 | 3 | 2 |
| The Guardian | 98 | L | 5 | 19 | 18 | 18 | 18 | 18 |
| The NYT | 40 | L | 3 | 1 | 8 | 7 | 9 | 12 |
| The WSJ | 41 | - | 1 | 6 | 5 | 10 | 12 | 7 |
| USA Today | 30 | C | 5 | 4 | 3 | 7 | 7 | 7 |

## 4 CONCLUSION AND FUTURE WORK

We presented POLUSA, a dataset of 0.9M online news articles covering policy topics. POLUSA aims to represent the news landscape as perceived by an average US news consumer between Jan. 2017 and Aug. 2019. To achieve this, we performed a series of preprocessing steps, including near-duplicate removal as well as balancing temporally and based on the popularity of outlets. Further, we assigned political leanings to outlets based on prior results. In contrast to previous datasets, POLUSA allows to analyze differences in reporting across the political spectrum, an essential step in, e.g., the study of media effects and causes of political partisanship. With almost a million of articles, the dataset allows utilizing data-intense deep-learning methods. For future work, we plan to enlarge POLUSA with further languages, and further balance the dataset, e.g., geographically or across the political spectrum.